# Privacy Risks in Health Big Data: A Systematic Literature Review

Zhang Si Yuan1, Manmeet Mahinderjit Singh 1*

1*School of Computer Science, University Sains Malaysia, Gelugor 11800, Penang, Malaysia;

syzhang@student.usm.my

* Correspondence: manmeet@usm.my

**ABSTRACT**

The digitization of health records has greatly improved the efficiency of the healthcare system and promoted the formulation of related research and policies. However, the widespread application of advanced technologies such as electronic health records, genomic data, and wearable devices in the field of health big data has also intensified the collection of personal sensitive data, bringing serious privacy and security issues. Based on a systematic literature review (SLR), this paper comprehensively outlines the key research in the field of health big data security. By analyzing existing research, this paper explores how cutting-edge technologies such as homomorphic encryption, blockchain, federated learning, and artificial immune systems can enhance data security while protecting personal privacy. This paper also points out the current challenges and proposes a future research framework in this key area.

## 1.0 Introduction

In the big data era, data is ubiquitous. As an important strategic resource, healthcare big data frequently involves the collection, use, and processing of sensitive personal information (such as genetic information[1], electronic health records[2], and wearable device data[3]) amid the emergence of new technologies, making personal privacy protection an increasingly prominent issue. In daily life, big data is everywhere, and even our every move on the internet is recorded. Healthcare big data is a new concept that has emerged in recent years alongside digital transformation and informatization, referring to health datasets that cannot be captured, managed, and processed by conventional software tools within an acceptable timeframe. Professional processing and reuse of this medical data has significant implications for health monitoring, disease prevention, and health trend analysis. However, new challenges have emerged: Who are you? Where are you? What is the health status of you and your family? Do you have any underlying conditions? When an individual's scattered data is aggregated, personal privacy may cease to exist. As aptly described by some: in the big data era, everyone becomes "transparent."

In 2023, a staggering 171 million patient records were breached in the United

States. This figure includes breaches reported to the Department of Health and Human Services under HIPAA, as well as incidents involving health data held by U.S. entities not covered by HIPAA. Breaking down the data further, there were 1,161 reports involving covered and non-covered entities in 2023, totaling 171,139,241 breached records. In comparison, there were 1,138 reports in 2022, totaling 59,664,152 breached records. In July 2020, GED match announced that its servers suffered a sophisticated attack, resulting in millions of user records being exposed within three hours. GED match hosted approximately 17 million publicly accessible genetic records because the platform allows users to upload their own data to find relatives or build family trees. On June 30, 2021, researchers discovered an unencrypted database belonging to New York company Get Health, which exposed over 61 million sensitive health data records collected from various wearable devices and applications, including users' names, birth dates, weight, height, gender, and GPS logs.

This paper conducts a systematic literature review (SLR) on privacy risks in health big data. The SLR was conducted according to the Kitchenham method [4] and 104 important relevant publications published between October 2013 and January 2024 were retrieved. Based on the previous systematic literature review (SLR), this study reviews privacy protection in health big data from two aspects. First, we categorize existing studies according to the big data life cycle and the latest privacy protection technologies, covering their research methods, limitations, and suggestions. Second, based on the findings of the literature review, we propose a privacy protection framework to better protect health big data from the increasing number of cyber-attacks. This study focuses on the following three research questions: **RQ1.** What privacy risks does health big data currently face? **RQ2.** What solutions can be used to address the risks and challenges in health big data? **RQ3.** What are the legal and regulatory studies on health big data in the existing literature?

The paper is organized as follows: Section 2 introduces relevant background concepts and explains the basic principles of SLR cameras; Section 3 elaborates on the first two phases (planning and implementation) of the Systematic Literature Review (SLR); Section 4 presents the third phase of the SLR, focusing on research findings and discoveries; Section 5 discusses the challenges encountered and relevant recommendations; Section 6 proposes a conceptual privacy protection framework; and Section 7 concludes the paper.

**2.0 Historical Background**

This section provides an overview of key concepts, including health big data, privacy risks, and existing protection mechanisms. It also introduces the technological and regulatory frameworks that underpin data privacy in this domain.

**2.0.1 Surveys on Health Big Data**

Health big data refers to the vast, diverse, and rapidly changing collections of data generated in the field of healthcare. These data come from various sources including electronic health records, medical devices, health monitoring tools, social media, and genomic data[5]. The characteristics of these data can be summarized by the "5V" model: Volume, Velocity, Variety, Veracity, and Value[6]. Specifically, Volume refers to the massive amounts of data such as patient records, medical imaging, and laboratory results. Velocity emphasizes the need for real-time or near-real-time processing capabilities due to the fast generation and processing of health data. Variety encompasses both structured and unstructured data. Veracity is crucial to ensure data accuracy and avoid erroneous medical decisions. Value indicates that insights and knowledge extracted from these data can improve patient care, optimize medical processes, and advance medical research, bringing significant social and economic benefits. Although health big data has wide applications like disease prediction, personalized medicine, and drug development, its management and analysis also face challenges such as privacy protection, data security, data standardization, and interoperability.

### 2.0.1.1 The definition of electronic health records.

Electronic health record (EHR) is used to create and manage health records of patients in a digital format[2]. It works like a central repository to store patients' health status information. Information such as medicines, medical treatment such as operations, historical information of past diagnoses, results of laboratory and radiology test and other health-related information of patients are securely and instantly available to patients and doctors[7].

### 2.0.1.2 The definition of genomic data.

Genomic data[1] refers to all genetic information obtained from an organism's genome through DNA sequencing technologies. It includes the sequence of base pairs in the DNA, the location and functional annotations of genes, genetic variations such as single nucleotide polymorphisms (SNPs) and insertions or deletions, as well as epigenetic data like DNA methylation and gene expression levels. Genomic data plays a crucial role in understanding gene functions, biological evolution, genetic variations, and disease mechanisms. Particularly in projects like the Human Genome Project, genomic data has been widely used to advance research in genetics, medicine, and bioinformatics.

### 2.0.1.3 The definition of wearable device data.

With the recent shift towards person (patient)-centered care and the widespread adoption of self-tracking technologies, wearable devices such as smart watches, armbands, and glasses are quickly becoming integral to people's daily lives. Most of these devices incorporate biometric sensors that can record and analyze health indicators, making the data they generate a valuable source for biomedical research. [8]A 2019 survey indicated that 38% of Americans currently use technologies like mobile apps or wearables to monitor their health data, with 28% having done so in the past. [9] Examples of person-generated health data (PGHD) include passively collected data from sensors, such as step counts, heart rate, and sleep quality; data actively entered by individuals like diet, stress levels, and quality of life; and social or financial information that, while not specifically health-related, could provide health insights. [10]Among the various types of PGHD, data generated by wearable devices are unique because they are passively, continuously, and objectively collected in free-living conditions, unlike other technologies that require manual data input (e.g., dietary tracking mobile apps). While existing wearable devices can typically only collect data from a few millimeters below the skin's surface, ultrasound can image the inside of the body through sound waves, just like sonar underwater, making medical imaging convenient and affordable. [3]Although existing wearable devices typically collect data from only a few millimeters beneath the skin, ultrasound can use sound waves to image the inside of the body, similar to underwater sonar, making medical imaging more convenient and cost-effective.[11-14]Thus, wearable device data generated by individuals are becoming an increasingly valuable resource for biomedical researchers, offering a more comprehensive picture of both individual and population health.

### 2.0.2 Threats to Privacy, Privacy Vulnerabilities, and Privacy Safeguards

**2.0.2.1 The definition of privacy Attack**

Privacy attacks[15]refer to the behavior of attackers to obtain, analyze and use personal information through various means to identify or infer the identity, behavior, preferences or other sensitive information of a specific individual. Such attacks usually use quasi-identifiers (such as age, gender, zip code, etc.) and other publicly available information in the data set to re-identify the real identity of the data subject through methods such as data linking, inference or collision, thereby invading personal privacy. Common types of privacy attacks include link attacks, membership attacks, inference attacks and homology attacks.

**2.0.2.2 The definition of privacy risk**

Privacy risk [16]refers to the probability that individuals may face issues during data processing, and the consequences if these issues occur. The scope of privacy risk includes but is not limited to technical measures lacking appropriate security protections, suffering from social media attacks, malware on mobile devices, unauthorized access by third parties, negligence due to incorrect configurations, use of outdated security software, being affected by social engineering, and the lack of encryption methods.

### 2.0.2.3 The definition of privacy protection

Privacy refers to an individual's right to decide whether information about him or her is released, while confidentiality is an assurance given by a data holder that they will not violate any individual's privacy by releasing data the individual desires to be private. Privacy prevents information about a person being shared, and confidentiality ensures data relating to that individual are shared only with authorized parties, while his or her identity is protected[17].

### 2.0.3 The Six Key Technologies

### 2.0.3.1 Homomorphic Encryption

The concept of homomorphic encryption was first proposed by Rivest and others in 1978[18], also known as privacy homomorphism. Fully homomorphic encryption has been hailed as the "Holy Grail" [19]of cryptography, as it has become a significant open problem in the field since its inception. Homomorphic encryption is a method of encryption that allows computations to be performed on encrypted data without requiring decryption. This means that operations on ciphertext will yield results that, when decrypted, correspond to the same results as if the computations were performed on the plaintext. Specifically, homomorphic encryption relates to a mapping in abstract algebra. For example, in an additive homomorphic encryption (HE) scheme, given encrypted messages $E(m_1)$ and $E(m_2)$ one can obtain $E(m_1 + m_2)$ using only $E(m_1)$ and $E(m_2)$, without knowing the original messages $m_1$ and $m_2$, here, $E(m)$ represents the encryption function, and $n$ is part of the public key. This means that multiplying the ciphertexts yields the encryption of the sum of the plaintexts[20], as formula (1) describes.

$$E(m_1 + m_2) = E(m_1).E(m_2)(mod\ n^2) \quad (1)$$

In 2009, Gentry developed the first fully homomorphic encryption algorithm[21], overcoming a problem that had puzzled researchers for over 30 years. This marked a significant advancement in both cryptography and computer science and initiated the era of fully homomorphic encryption research. Numerous FHE schemes have been developed since, aiming to improve efficiency, reduce computational overhead, and enhance security. The Evolution of Homomorphic Encryption is shown in Figure 2-1.

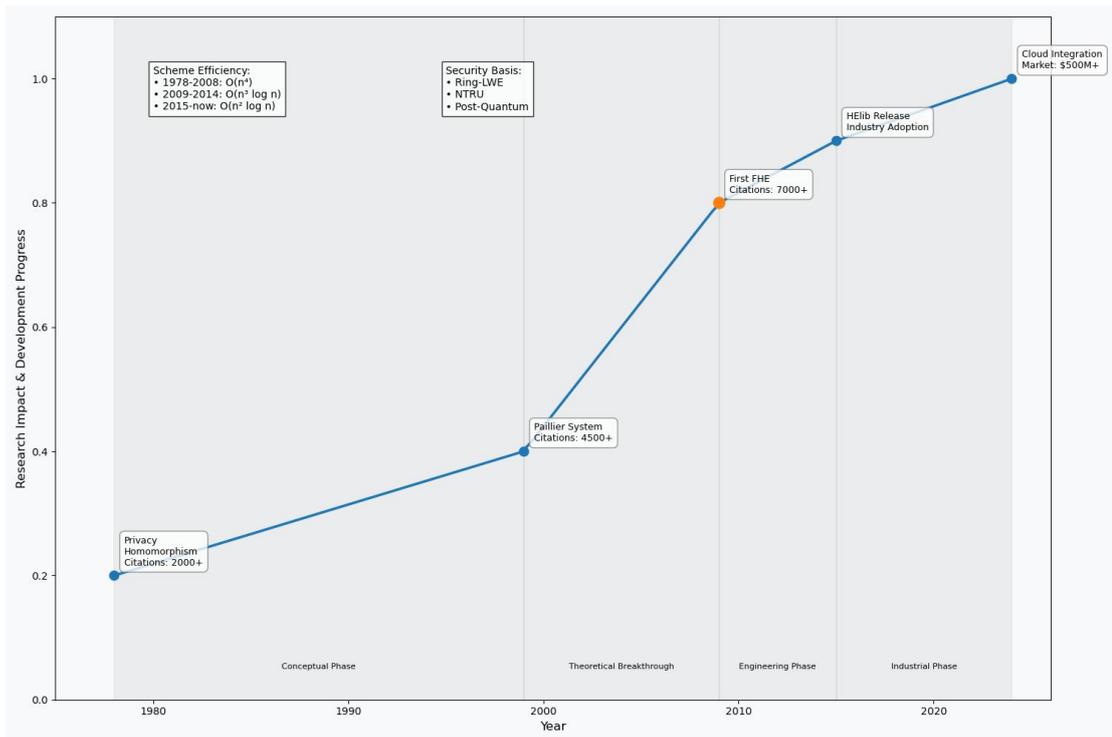

**Figure 2-1 Evolution of Homomorphic Encryption (1978-2024)**

### 2.0.3.2 Zero-Knowledge Proof

Zero-Knowledge Proof (ZKP) is a cryptographic technique that allows a prover to prove to a verifier that a certain statement is correct without providing any additional information beyond the correctness of the statement. The prover can prove knowledge of the preimage of a hash value, or knowledge of a member of a Merkle tree with a known Merkle root, without revealing any information about the witness[22, 23].

Feige et al. [24] formally defined an interactive zero-knowledge proof system, consisting of a pair of polynomial-time probabilistic Turing machines (a prover and a verifier), which satisfies the following properties for a polynomial-time predicate $P(I, S)$:

1. For all inputs I that satisfy $P(I, S)$, the interaction between the honest prover and verifier succeeds with overwhelming probability.

2. There exists a machine $M$ such that for all provers $A$, random tape $RA$, and sufficiently large input $I$, if the execution of $A$ and verifier $B$ on input $I$ succeeds with non-negligible probability, then the output produced by $M$ at the end of the execution satisfies predicate $P$ with overwhelming probability.

Zero-knowledge proofs enable the prover to convince the verifier that they have correctly performed a computation on a set of inputs without revealing some of the inputs, and the information required for verification is much smaller than the computation itself [23]. Zero-knowledge proofs have been proven to effectively solve many problems[22] and have wide applications in the blockchain field, enabling important features such as privacy protection[23].

**2.0.3.3 Post-quantum Cryptography**

Post-quantum cryptography is the study of cryptographic algorithms and protocols that remain secure in the era of quantum computers. Since quantum computers can efficiently solve certain mathematical problems relied upon by classical cryptography, such as integer factorization and discrete logarithm problems, the goal of post-quantum cryptography is to develop new cryptographic algorithms that can resist quantum attacks [25].

**2.0.3.4 Blockchain Technology**

**2.0.3.4.1 The origin and development of Bitcoin**

When discussing blockchain technology, people often first think of Bitcoin, as blockchain was originally developed as the underlying framework for Bitcoin. Therefore, before exploring blockchain technology further, let's first briefly understand its origins—Bitcoin.

In November 2008, a person using the pseudonym Satoshi Nakamoto published a paper titled "Bitcoin: A Peer-to-Peer Electronic Cash System" on the cryptography forum metzdowd.com. The paper provides a detailed explanation of the key technological concepts of decentralized digital currencies. The system aims to achieve peer-to-peer transactions through blockchain technology, eliminating reliance on central authorities and securely validating and recording all transactions[26]. On January 3, 2009, Satoshi Nakamoto launched the Bitcoin system and mined the first block, known as the Genesis Block, marking the creation of the first 50 bitcoins[27]. Since the birth of Bitcoin, extensive research and discussion have been conducted in both academia and industry. Many studies have focused on improving Bitcoin's scalability, security, and privacy. For example, Gervais et al. [28] used a quantitative framework to analyze the security of PoW blockchains and found that Bitcoin is more secure than Ethereum.

**2.0.3.4.2 The definition of blockchain**

Blockchain technology is essentially a decentralized database and serves as the core technology and infrastructure of Bitcoin. It represents a new application model for various computer technologies such as distributed data storage, peer-to-peer transmission, consensus mechanisms, and cryptographic algorithms[29]. In a narrow sense, blockchain is a chain-like data structure that sequentially combines data blocks in chronological order and is a distributed ledger that is difficult to tamper with or forge, safeguarded by cryptography. In a broader sense, blockchain technology is a new distributed infrastructure and computing paradigm that utilizes blockchain data structures to verify and store data, employs distributed node consensus algorithms to generate and update data, uses cryptography to ensure the security of data transmission and access, and relies on smart contracts composed of automated script codes to program and manage data[30]. The history of blockchain is shown in Figure 2-2.

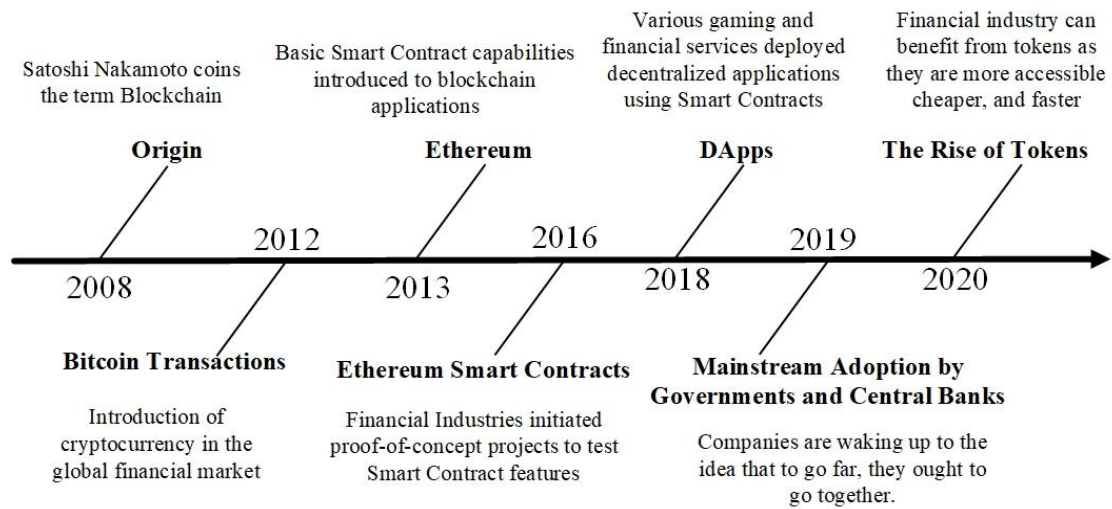

Figure 2-2 History of Blockchain

### 2.0.3.5 Federated Learning

#### 2.0.3.5.1 Definition

Federated learning [31]aims to develop a model based on distributed datasets without sharing the raw data. It involves two key processes: model training and model inference. During the training phase, parties can exchange model-related information (possibly in encrypted form), but not the data itself. This exchange ensures that no protected private information from any site is exposed. The trained federated learning model can be deployed at each participating entity in the system or shared among multiple parties.

#### 2.0.3.5.2 Example Application Scenarios

Figure 2-3 illustrates an example of a federated learning architecture that includes a coordinator. In this scenario, the coordinator is an aggregation server (also known as a parameter server) that can distribute the initial model to Hospitals A through C. Hospitals A–C individually train the model using their own datasets and send the updated model weights back to the aggregation server. The aggregation server then aggregates the model updates received from each hospital (for example, using the Federated Averaging algorithm[32]) and sends the aggregated model updates back to the hospitals. This process repeats until the model converges, reaches the maximum number of iterations, or the maximum training time is achieved. In this architecture, the participants' original data never leaves their own devices. This method not only protects user privacy and data security but also reduces the communication overhead associated with transmitting raw data. Additionally, the aggregation server and participants can use encryption methods (e.g., homomorphic encryption[33, 34]) to prevent model information leakage [35].

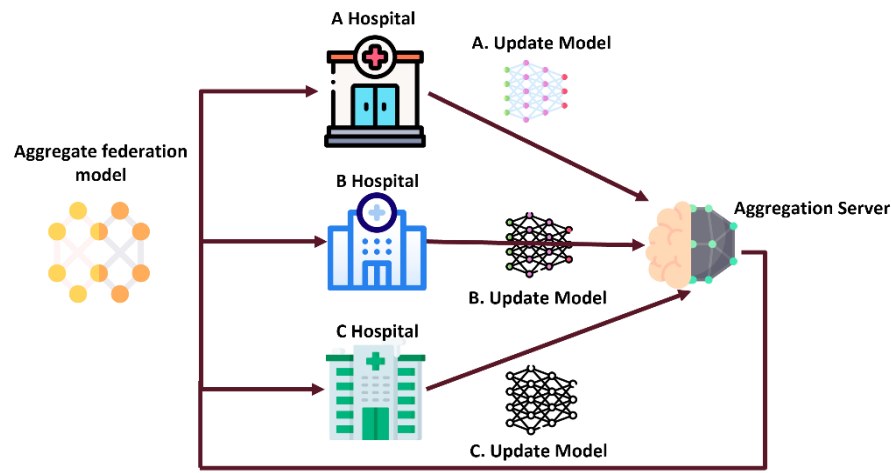

**Figure 2-3 Example of a federated learning system: client-server architecture**

**2.0.3.6 Human Immune System**

**2.0.3.6.1 Definition**

Theoretical immunology and artificial immune systems are two distinct research areas, both related to the study of the human immune system. Theoretical immunology, including computational immunology, focuses on explaining immune phenomena and solving immunological or medical issues, with a long-standing historical foundation. In contrast, artificial immune systems, which approach immune mechanisms and properties from an engineering and scientific perspective, aim to develop innovative solutions for various problems, including medical challenges. This field is also known by different terms, such as artificial immune systems, immune-based systems, and immune computing, though these terms are often used interchangeably in the literature.

Consequently, artificial immune systems have acquired multiple definitions. An artificial immune system can be regarded as a framework based on the principles of the human immune system, employing data processing, classification, representation, and reasoning strategies rooted in biological paradigms[36]. It also represents an intelligent method inspired by biological immune systems to address real-world problems[37], and a computational system modeled after the methods of natural immune systems[38]. In their monograph on the subject, De Castro and Timmis describe artificial immune systems as adaptive systems developed by leveraging immune system mechanisms and insights from theoretical immunology[39]. Internationally, the term " Immune Computing " is also used to encapsulate the main concepts of this field[40]. This article continues to adopt the more traditional concept of artificial immune systems to maintain the openness of research ideas.

### 2.0.3.6.2 The intelligent model of the immune system

An intelligent model of the immune system is presented, as illustrated in Figure 2-4. Immune cognition involves interactions among various elements within the multi-layered immune system, including a complex network of molecules, cells, and tissues. This network supports learning, recognition, and the formation of immune memory. Recognition, a primary and critical step, involves perceiving danger signals and distinguishing between self and non-self, while also learning and memorizing non-self-elements. Through repeated immune responses, the learning and memory processes are continuously refined. These cognitive functions of recognition, memory, and learning are generated through the dynamic interactions within this immune network.

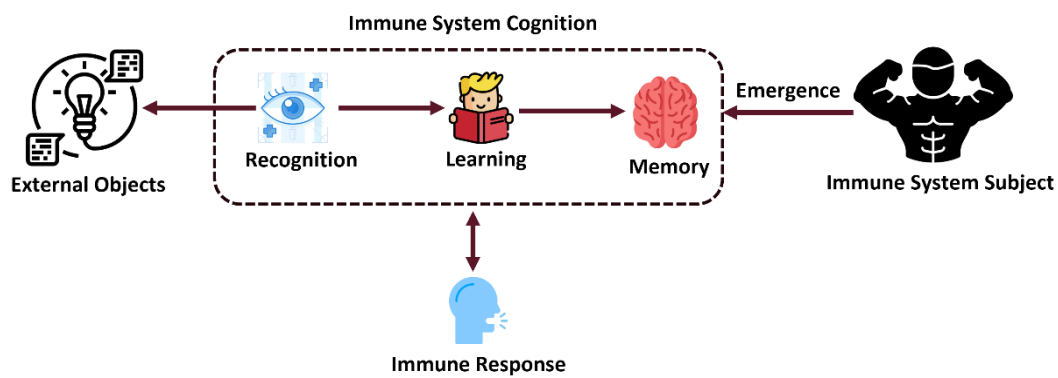

**Figure 2-4 Intelligent Model of the Immune System**

### 2.0.4 Relevant Laws, Regulations, and Ethical Considerations

In the process of collecting and analyzing health big data, adhering to relevant laws, regulations, and ethical guidelines is crucial to ensure the legality and ethicality of research[41]. The General Data Protection Regulation (GDPR), officially implemented in the European Union in 2018, has established strict standards for personal data protection, emphasizing respect for individual privacy and transparency in data processing[42]. According to the GDPR's requirements, the collection and processing of personal health data must obtain explicit consent and adopt appropriate technical and organizational measures to ensure data security [43]. The Health Insurance Portability and Accountability Act (HIPAA) in the United States provides specific regulations to protect medical information [44]. HIPAA's Privacy Rule and Security Rule require medical institutions to implement specific measures to safeguard personal health information, including controlling data access, ensuring transmission security, and training staff [45]. Violations of HIPAA regulations may lead to severe legal and economic consequences, including hefty fines and criminal penalties [46].

Ethical guidelines also play a critical role in guiding research practices[47]. The Declaration of Helsinki, as the international ethical benchmark for medical research, emphasizes respect for subjects, the importance of obtaining informed consent, and balancing risks and benefits [48]. The Belmont Report further proposes the three ethical principles of respect for persons, beneficence, and justice, providing guidance for research involving human subjects[49]. With the widespread application of big data and artificial intelligence in the medical field, traditional legal and ethical frameworks face new challenges [50]. For example, machine learning models may inadvertently leak personal sensitive information, increasing privacy risks [51]. Therefore, researchers need to strike a balance between technological innovation and ethical norms, adhere to the principle of minimization, and avoid excessive collection and use of personal data[52]. The international community is calling for the establishment of new ethical frameworks to address the challenges of the digital age [53]. The European Commission's "Ethics Guidelines for Trustworthy AI" emphasizes the importance of transparency, accountability, and human oversight, aiming to ensure that the development of AI technology conforms to ethical standards [54]. In summary, complying with relevant laws, regulations, and ethical guidelines is essential for safeguarding participants' rights, ensuring research credibility, and promoting sustainable scientific development [55].

**3.0 Planning and Implementation of the SLR**

To identify and classify the literature related to privacy risks in health big data, we adopted a scientific, comprehensive, and reproducible approach to conduct a Systematic Literature Review (SLR). According to Kitchenham[4], an SLR comprises three phases: planning, conducting, and reporting. This section presents the planning and conducting phases, while the Results and Findings phase is detailed in Section 4.

**3.0.1 Planning the SLR**

The planning process for the systematic literature review (SLR) involved the following activities:
(1) Formulating research questions to define the scope of the search;
(2) Defining and refining search strings.
(3) Selecting appropriate data sources;
(4) Establishing clear inclusion criteria.

**3.0.1.1 Formulation of the research questions.**

The main objective of our SLR is to assess the current research status on privacy risks in health big data and understand how to mitigate these privacy risks. To achieve this objective, we developed the following research questions:
RQ1: What are the current privacy risks associated with health big data?

RQ2: What solutions are available to address the risks and challenges in health big data?

RQ3: What are the legal and regulatory studies on health big data in the existing literature?

**RQ1** identified important privacy risks in health big data, such as data breaches, unauthorized access, re-identification, and misuse of personal information. **RQ2** explored technical solutions, including blockchain, encryption, anonymization, access control, cloud computing privacy protection, and federated learning, to mitigate privacy risks in health data. **RQ3** reviewed the literature on legal frameworks (such as GDPR and HIPAA) and their role in addressing privacy issues in health big data.

**3.0.1.2 Defining and refining search strings.**

We developed a 12-query search strategy to investigate privacy issues in health-related big data. This strategy was based on two foundational pillars: (i) a preliminary review of seminal literature identifying key privacy vulnerabilities, and (ii) our collective expertise in this area.

Initially, "privacy risks" was employed as an overarching term to capture a wide range of relevant studies. To further refine the search, we incorporated terms such as "data breaches," "data sharing," and "health information," all closely linked to privacy challenges. The rationale for these choices is as follows:

- **Data breaches**: These involve unauthorized access or disclosure of health data, posing significant threats to privacy.
- **Data sharing**: The exchange of health data between entities or systems raises concerns about the security of shared information.
- **Health information**: As a critical element of health data, personal health information is vulnerable to privacy breaches when mishandled or exposed.

Using these criteria, we constructed search queries such as:

1. Articles intersecting "health big data" with "data breaches," "data sharing," or "health information."
2. Studies examining "privacy protection" alongside these terms.
3. Research exploring the relationship between "privacy attack" and the specific terms.
4. Articles addressing "data privacy" in conjunction with these terms.

Each query was designed to focus on a specific aspect of privacy concerns, resulting in a comprehensive set of 12 search strings, including combinations such as "health big data and data breaches" and "health big data and health information."

**3.0.1.3 Selection of Data Sources**

This systematic literature review (SLR) used one data source: scientific digital libraries The libraries included ACM Digital Library, IEEE Xplore, Springer Link, ScienceDirect, Web of Science, Scopus, PubMed, and Google Scholar. We found that search engines in these libraries required customized queries due to differences in syntax and search options, such as searching by title, abstract, or full text.

All searches spanned entire databases, reflecting the multidisciplinary nature of privacy risks in health big data, covering fields like healthcare, data security, IT, and law. We used specific syntax, placing multi-word terms (e.g., "privacy risks") in quotes and combining search terms with "AND," except in ACM Digital Library, which required "+" instead. These sources were chosen to ensure comprehensive coverage of the literature.

**3.0.1.4 Definition of inclusion criteria.**

To address privacy risks in health big data, as suggested by sources such as [56-59], inclusion criteria should be defined to ensure the fair selection of relevant publications. Studies that meet the following criteria will be included, as these standards help to identify research that provides significant and high-quality contributions:

IN1: This study explores issues related to privacy risks and focuses on the protection of privacy in health big data.
IN2: The publication has undergone a peer-review process.
IN3: If the same study appears in multiple publications with partial results, the one providing the most comprehensive findings is selected.
IN4: The publication is written in English.

**3.0.2 Implementation of the SLR**

After the planning phase, a systematic literature review (SLR) was conducted between October 2013 and January 2024, with a search process divided into two phases:
As illustrated in Figure 3-1, the selection process was divided into three stages, as follows:
**(1) Phase 1 - Digital Library Search**: Following the query method described in Section 3.1.2, a search was conducted in each of the digital libraries listed in Section 3.1.3. The search covered journal articles published between 2013 and 2023, totaling 5,187 publications. After removing duplicates, 5,101 publications remained. The remaining papers were further reviewed by examining their introductions and conclusions, and papers from journals not classified as Q1 or Q2 were excluded. This process reduced the dataset to 88 papers.
**(2) Phase 2 - Backward snowball search[60]:** The references and citations of the publications identified in Phase 1 were analyzed to identify additional relevant studies. In this phase, 16 more papers were added, bringing the total number of relevant publications to 104.

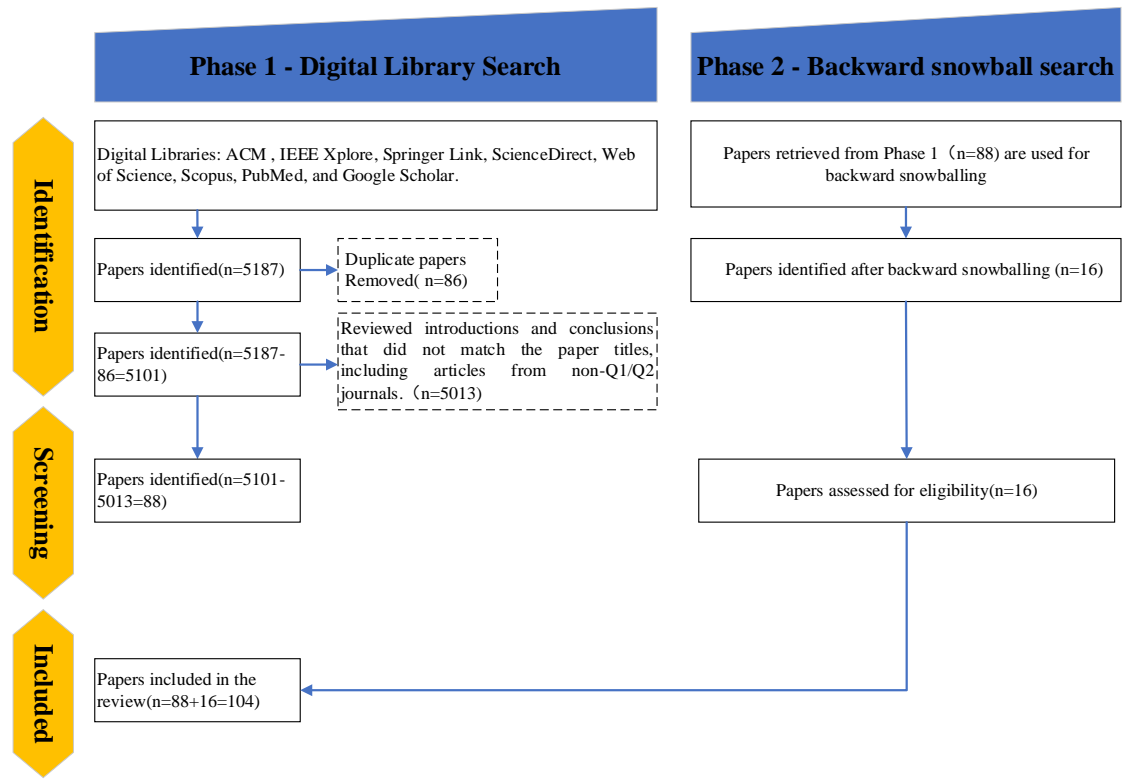

**Fig. 3-1. Flowchart summarizing the selection of publications in the two-stage search**

**4.0 Results and Findings from the Research Questions**

This section reports the analysis of the SLR results framed along the four research questions presented in Section 3.1.1.

### 4.0.1 What are the current privacy risks associated with health big data? (RQ1)

For the first question, this study selected 21 articles for in-depth analysis. Through the analysis and comparison of these 21 papers, it can be found that there are some commonalities and individualities in the research on privacy and security issues of medical big data. In terms of commonalities, these papers all believe that while medical big data brings huge opportunities to the medical industry, it also faces severe privacy and security challenges. Privacy protection has become a problem that must be faced and solved in the development of medical big data. It is widely believed that there are varying degrees of privacy leakage risks in data collection, storage, use and other links, and effective privacy protection measures need to be taken. Most papers focus on the privacy risks brought by attacks such as sensitive information leakage, data abuse, and re-identification[61-69]. Most papers mainly analyze privacy risks from a technical perspective, and discuss the limitations of technical mechanisms such as encryption, desensitization, and access control[61, 63-66, 68, 70-72]. Most papers focus on analyzing the causes and consequences of risks in the description of privacy risks, and the excavation of the intrinsic mechanisms of risks is not enough[62, 63, 66, 67, 69-80]. In terms of individuality, although everyone is aware of the importance of privacy and security, the focus of risk concerns varies. Some focus on sensitive data collection[70, 80], some focus on storage risks[71] [72, 79], and others pay special attention to privacy issues in secondary use[67, 81]. In terms of the breadth of data lifecycle coverage, most papers only analyze the privacy risks of certain links [61-65, 67-73, 75, 76, 79, 80], while[66, 74, 78, 81] attempt to systematically examine privacy issues from a full lifecycle perspective. Although privacy attacks are discussed, the depth of attack methods and techniques varies. Most papers stay at the conceptual level, while [64, 65, 79, 81] conduct more in-depth analysis of certain attack techniques. In the analysis of the causes of privacy risks,[76, 77] cut in from the perspective of insufficient user privacy awareness, [78] focuses on the problem of heterogeneous data fusion, and [81]analyzes emerging risks such as data pollution and model attacks, all of which reflect the unique perspectives of the authors. In terms of research ideas, [61, 62, 64, 65, 68, 70] follow the idea of "risk identification-threat analysis-protection measures", while [76, 77, 81] also attempt to discuss solutions from non-technical perspectives such as institutions and ethics, reflecting the differences in the authors' ideas for analyzing problems. In terms of systematic analysis of the problem, [66, 74, 78, 81] strive to analyze the problem with a global perspective, but most papers still mainly focus on a certain aspect of privacy and security, and the analysis of the problem is not comprehensive and systematic enough. It can be seen that the existing literature has consensus on the basic viewpoints in the research on privacy and security issues of medical big data, but there are still differences in the focus of attention, research breadth, and analysis depth.

To ensure the rigor and accuracy of the classification standards, this issue has been categorized according to the following medical big data lifecycle stages, as shown in Table 4-1.

**Table 4-1 Taxonomy Classification of Privacy Risks and Attacks in Healthcare Big Data Lifecycle**

| Data Lifecycle Stage | Privacy Risk Category | Detailed Description | Privacy Attack Methods | System Vulnerabilities | References |
|---|---|---|---|---|---|
| **Data Collection** | Sensitive Information Leakage | The collection of medical big data inevitably involves patients' sensitive information, including personally identifiable information (PII). If protection measures are inadequate, it can easily lead to unauthorized leakage of this private data. | Data breach attacks, information gathering attacks, re-identification attacks | Lack of encryption for sensitive data, insufficient de-identification methods, data correlation risks | [66, 68, 70, 73, 75, 80] |
| | Unauthorized Access | Collecting, accessing, or misusing user data without explicit authorization, and the scope of data use exceeds necessary limits, violating the data minimization principle and constituting an invasion of personal privacy. | Privilege escalation, data mining, information inference | Weak data access control mechanisms, misconfigured permissions, lack of transparency in data collection process | [62, 63, 74, 81] |
| | Lack of Privacy Awareness | Users have insufficient awareness of the potential privacy risks after data is collected. Especially in innovative application scenarios such as intelligent healthcare, due to a lack of necessary transparency, users cannot effectively exercise their right to be informed. | | Insufficient transparency of data usage information, inadequate privacy risk notification, users' lack of privacy awareness | [76, 77] |
| | Data Quality Issues | The collected data may have inherent biases, affecting the fairness of analysis results. | Data pollution, data distortion | Data bias leads to unfair analysis, unclear data ownership, lack of | [77, 81] |
| | | Moreover, in the big data environment, data ownership boundaries are blurred, and data control rights can be easily abused. | | supervision on third-party access | |
| | Data Integration | Introducing social network data for auxiliary analysis may cause privacy leakage. Correlating environmental data with health data may infer an individual's living environment and health conditions. | Data correlation, data aggregation | Cross-correlation of data with external social media data, lack of privacy protection in heterogeneous data source integration and analysis | [78] |
| **Data Storage** | Data Leakage | Medical big data is highly susceptible to network attacks during storage, resulting in large-scale leakage of private data. Vulnerabilities in the storage system itself may also be maliciously exploited. The integrated storage of heterogeneous data sources is more likely to cause privacy leakage. | APT attacks, supply chain attacks, data theft | Improper data encryption, insufficient security of storage systems, security vulnerabilities in systems, improper integration of heterogeneous data | [61, 63, 70-74] [66, 78, 79] [78] |
| | Integrity Threats | Stored data faces the risk of being tampered with or destroyed, affecting data integrity. Existing encryption technologies have insufficient performance when processing massive heterogeneous data. | Data tampering, cryptanalysis | Lack of integrity verification in storage systems, inefficient application of encryption algorithms to large-scale data | [63-66, 72, 74] |
| | Re-identification Risks | Using public data or other external information, attackers can perform re-identification attacks on de-identified data, leading to risks such as identity theft. | Linkage attacks, background knowledge attacks | Imperfect data de-identification methods, high risk of external data correlation. Weak identity authentication system, lack of multi-factor authentication. | [64, 67-69, 72, 81] |
| | Security and | Big data storage and transmission require strict | - | Inadequate data encryption and de- | [77, 78] |
| | Availability Contradiction | security protection, but at the same time, data availability should not be excessively affected. Data heterogeneity makes privacy protection more complex. | | identification mechanisms. Improper integration of heterogeneous data sources, insufficient privacy protection. | |
| **Data Analysis and Utilization** | Privacy Feature Inference | Big data analysis algorithms may disclose user identities, sensitive attributes, or other private information through data re-identification, feature inference, and other methods. | Machine learning attacks, shadow attacks | Incomplete data de-identification, inference risks in machine learning models, lack of protection in data correlation and integration | [61, 64, 65, 68, 69, 73-75, 81] |
| | Data Misuse | The purpose of data use may deviate from the original intention. Data users may discriminate against individuals based on analysis results. Biases in statistical analysis can also exacerbate unfairness. | Information discrimination, algorithmic bias | Lack of supervision on data usage behavior, data use exceeding agreed scope. Selection bias in training data, discrimination risks in algorithms. | [62, 66, 67, 69, 74, 76, 79], |
| | Model Side-Channel Information Leakage | Malicious parties can extract private data from machine learning models through model inversion, membership inference, and other means. Model outputs may also expose sensitive information. | Reverse engineering of models, privacy tracking | Insufficient protection of machine learning models, ease of extracting training data. Insufficient sensitivity analysis of model outputs. | [64, 65, 81] |
| | Opaque Data Processing | The purpose of data use is not clearly stated, and data subjects have difficulty knowing the actual use of their data. The data processing flow is opaque, and data quality issues may lead to erroneous decisions and privacy violations. | | Low transparency of data processing information, easy deviation in data interpretation, poor data quality control | [62, 66, 69] [78] |
| | Real-time Processing Challenges | The demand for real-time data generation and analysis is growing, and privacy protection | | Rapid data generation and processing, privacy protection mechanisms | [78] |

| | | measures urgently need to be improved to address new complexities. | | struggle to keep up in real-time | |
|---|---|---|---|---|---|
| **Data Destruction** | Incomplete Destruction | During the data destruction process, removal is incomplete, and residual sensitive data may be recovered. Improper disposal of obsolete storage devices can also lead to data leakage. | Data reconstruction, forensic analysis | Imperfect data deletion mechanisms, easy recovery of residual information. Lack of security measures in the disposal process of storage devices. | [63, 71, 74, 80, 81] |
| | Data Difficult to Erase | Limited by technical capabilities, management systems, or data's own characteristics (such as the immutability of blockchains), some data is difficult to completely destroy. | Distributed storage attacks | Existing technologies and systems cannot yet ensure the 'right to be forgotten' for sensitive data. New technologies like blockchains highlight the difficulty of deleting data. | [63, 74, 75, 81] |

Based on the analysis of 21 literatures, the bar chart "Distribution of Privacy Risk Categories in Data Lifecycle Stages of Big Data Analytics" illustrates the distribution of privacy risk categories across four stages of the data lifecycle (data collection, storage, use, and disposal). The data use stage has the highest number of risk categories at 32, accounting for 36.4%, indicating that this is the most complex and critical stage for privacy protection. This is followed by the data collection stage with 28 risk categories, representing 31.8%, reflecting the need to fully consider privacy protection issues from the initial data gathering phase. The data storage stage contains 23 risk categories, accounting for 26.1%, which, although fewer than the collection and use stages, still requires significant attention. The data disposal stage shows notably fewer risk categories with only 5, representing 5.7%, possibly due to limited research in this area or the relatively singular nature of risks at this stage. This distribution reveals that the focus of privacy protection in big data analytics should be primarily on the data use and collection stages, while also highlighting the need not to overlook privacy protection research in the data disposal stage. as shown in figure 4-1.

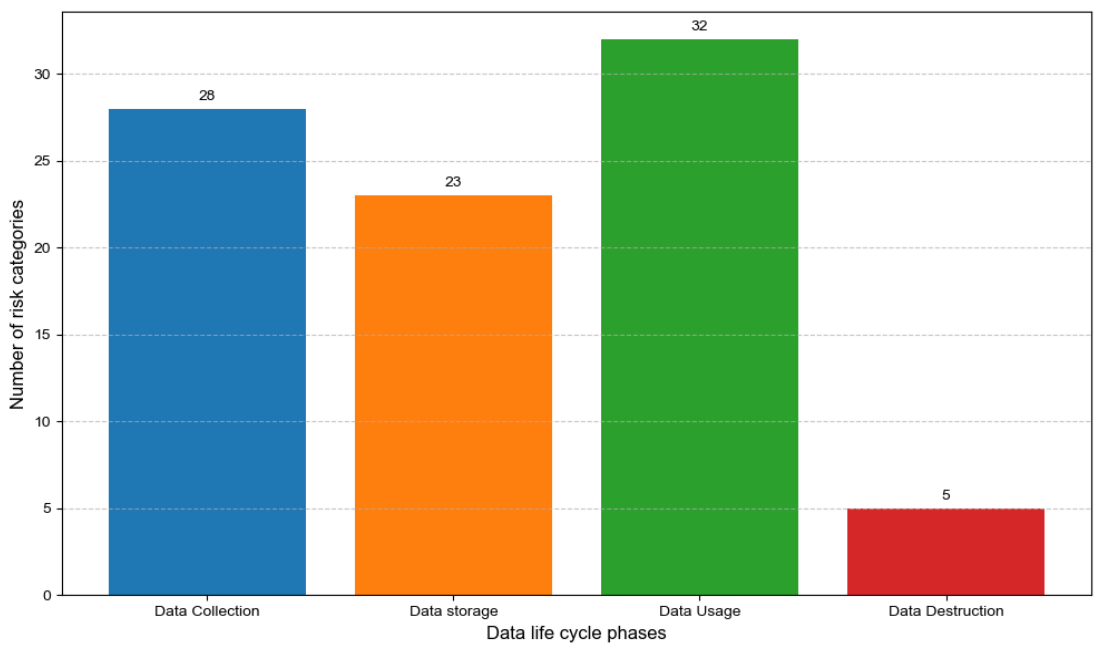

**Fig. 4-1. Frequency of occurrence of privacy risk categories in data lifecycle stages**

### 4.0.2 What solutions are available to address the risks and challenges in health big data？(RQ2)

For the second question, this study selected 52 articles for in-depth analysis. It can be seen that, although there are various privacy protection technologies in the medical and healthcare field, they all aim to maximize the protection of patient privacy during data sharing and analysis. These technologies can be roughly divided into several categories: blockchain, access control, cloud computing, data desensitization, cryptography, and federated learning.

Blockchain technology has been widely used in the medical and health field due to its decentralization, tamper-proofing, and traceability. Literature [82] proposes a blockchain-based secure storage mechanism for spatial-temporal big data, which supports efficient batch query verification while ensuring data security and traceability through on-chain and off-chain collaborative storage and updatable sub vector commitments. Literature [26] discusses the potential of Bitcoin's decentralization in solving medical data sharing and auditing from the perspective of electronic cash. Literature[83, 84] further explores the combination of blockchain with other technologies such as attribute-based encryption and proxy re-encryption to achieve fine-grained access control and secure sharing of medical data. However, blockchain technology also has some limitations, such as storage capacity and computational performance constraints, as well as high communication overhead [85, 86]. Literature [87] proposes a hierarchical data sharing framework based on blockchain, which achieves fine-grained permission management while improving storage efficiency through on-chain and off-chain hybrid storage, dynamic access control, and ciphertext splitting mechanisms. Literature[88] further explores the role of blockchain in attribute-based encryption key management, using smart contracts to achieve traceable and revocable key management. These two papers expand the application scenarios and implementation methods of blockchain technology in medical data sharing and key management.

Access control is an important means of protecting medical data privacy. Literature[89-93] achieves fine-grained access control based on attribute-based encryption (ABE), allowing medical data owners to flexibly customize access policies according to user attributes. Literature[94] provides an adaptive access decision solution based on fuzzy trust evaluation. For resource-constrained environments such as the Internet of Things, literature[95, 96] designs lightweight access control mechanisms. Although access control is intuitive and effective, when the number of attributes is large, the computational overhead is high, revocation is difficult, and there is a heavy key management burden.

Cloud computing provides powerful computing power for medical data storage and processing, but also introduces new privacy challenges. Cloud computing technology is widely used in the field of medical and health privacy protection, with 40 out of 52 papers involved, accounting for as high as 76.9%[25, 85, 92, 97-122]. Among them, literature [102, 104, 114, 116] uses homomorphic encryption to achieve privacy-preserving cloud data storage and computation. Literature[115, 121] introduces trusted hardware such as Intel SGX to protect the confidentiality of data and computation. Literature [101, 103, 108]explores hybrid cloud and data partitioning strategies to balance local and cloud storage and computation. Literature[100, 110] combines proxy re-encryption and attribute-based encryption to achieve secure data sharing in a cloud environment. In response to the potential threats posed by quantum computing, literature[25, 120-122] focuses on the application of post-quantum cryptography in cloud storage. However, how to ensure the security of data in the face of internal and external attacks, and how to balance the storage and computational overhead of local and cloud environments, remain urgent problems to be solved. Literature [105] proposes a fine-grained auditing method for privacy leakage problems in cloud environments. This method uses trusted hardware-assisted dynamic taint analysis technology to track and trace the flow of privacy data.

Data desensitization achieves privacy protection through data transformation, perturbation and other techniques. Literature[123, 124] uses traditional models such as k-anonymity and l-diversity to achieve data publishing privacy; literature[103, 119] uses differential privacy theory to guide data noise addition; literature[125, 126] generates synthetic data to replace the original data for analysis. Data desensitization is intuitive and effective, but it also inevitably loses the authenticity and accuracy of the data, making it difficult to achieve both data utility and privacy. Literature [110], on the other hand, designs a privacy quantification method based on fuzzy logic and regression analysis at the algorithm level, which evaluates user historical behavior in multiple dimensions and adaptively adjusts the strength of privacy protection.

Cryptographic techniques are an important cornerstone of privacy protection. Technologies such as homomorphic encryption[97, 98, 102, 106, 107, 109, 113, 114, 116], secure multi-party computation[97, 114, 118], and zero-knowledge proofs[84, 85] are widely used in medical data privacy protection. In particular, literature [90] introduces function encryption into fine-grained access control of electronic medical records. However, these cryptographic techniques also generally have limitations such as low computational efficiency, complex key management, and high communication overhead. Literature [117]designs an efficient ECG steganography for privacy leakage during medical data transmission. This method uses FWHT transform to construct a high-capacity steganographic space to protect data transmission security. Literature [122]starts from the physical layer and uses quantum encryption technology to protect the security of medical data transmission and storage, effectively resisting quantum computing attacks. These two papers provide new ideas for medical privacy protection from the perspective of data transmission and physical layer security.

Federated learning achieves machine learning while protecting data privacy through a distributed learning paradigm of local training and parameter aggregation. Literature[99, 127] combines blockchain and federated learning to achieve secure sharing and privacy-preserving analysis of medical data. Literature[111, 112] discusses the privacy advantages and potential of federated learning in the medical and health field. Literature [118, 119]further enhances the privacy of federated learning by introducing differential privacy and homomorphic encryption to protect gradient information. Literature [98] combines federated learning and blockchain to achieve privacy-preserving distributed collaborative learning in the medical image segmentation scenario. Literature[109] further optimizes the efficiency of federated learning by introducing offline/online aggregation and pruning techniques to reduce communication and computational overhead. These two papers demonstrate the potential of federated learning techniques in medical AI applications. However, federated learning also faces challenges such as performance bottlenecks, insufficient model robustness, and difficulty in modeling boundary data.

  The following is a detailed classification of the technologies involved in these papers and a summary of their advantages and disadvantages: as shown in table 4-2.

**Table 4-2 Comparative Table of Privacy Protection Techniques for Healthcare Big Data: Methods, Advantages, and Limitations**

| Technology Category | Method | Advantages | Disadvantages | References |
|---|---|---|---|---|
| Blockchain | Store encrypted hash or verification code on blockchain, control access and verification with smart contracts | Decentralized, tamper-proof, traceable, ensures data integrity and legitimacy, smart contracts improve efficiency | Limited storage and computation performance, high communication overhead for frequent interactions, lack of privacy protection | [26, 82-86, 128] |
| | Combine blockchain with homomorphic encryption, secure multi-party computation, etc., to protect privacy | Enhance privacy protection on top of blockchain security, support data usage and computation | High computation overhead of homomorphic encryption, low efficiency of multi-party collaboration, high system complexity | [84, 97, 98, 102] |
| | Combine blockchain with proxy re-encryption, attribute-based encryption, etc., to implement access control | Fine-grained dynamic authorization, flexible ciphertext conversion and attribute revocation, traceable access verification on blockchain | Complex key management, high overhead when the number of attributes is large, limited blockchain performance | [83, 88, 89, 100, 128] |
| | Hybrid application architecture of different types of blockchains, such as consortium chains and sidechains | Public chain as trust foundation, controllable permission in consortium chain, sidechains extend storage and computation capabilities | Difficult to maintain interoperability and consistency across chains, trade-off between security and performance | [83, 86, 87, 103] |

| | Federated learning or privacy-preserving machine learning based on blockchain | Verification and incentive in distributed collaborative learning process, protect model and parameter privacy | Consensus and encryption overhead affect learning efficiency, participant dropout leads to performance degradation | [98, 99, 127] |
|---|---|---|---|---|
| **Access Control** | Fine-grained access control based on attribute-based encryption (ABE) | Flexible attribute-based authorization policy, support attribute revocation, strong attack resistance | High encryption and decryption overhead when the number of attributes is large, high attribute revocation overhead, complex key management | [89-93] |
| | Traditional autonomous access control models based on roles or users | Simple and intuitive, easy to manage and implement | Poor flexibility, difficult to cope with dynamically changing requirements | [94] |
| | Privacy-preserving access based on homomorphic encryption and secure multi-party computation | Implement access control without decryption, protect data privacy | High computation overhead, low verification and authorization efficiency | [90, 114] |
| | Decentralized verifiable access control integrated with blockchain | Blockchain stores authorization records, tamper-proof, traceable | Authorization verification efficiency depends on blockchain performance, insufficient privacy protection | [83, 84, 88, 89] |
| | Efficient access control mechanisms for lightweight devices | Reduce computation overhead for resource-constrained devices, fast response | Reduced security, lack of fine-grained control and flexible revocation | [95, 96] |
| | Homomorphic encryption and secure multi-party computation | Privacy-preserving data usage and computation without decryption, multi-party collaborative learning | High encryption and decryption computation overhead, frequent multi-party interactions, low communication efficiency | [102, 104, 114, 116] |

| | Trusted hardware and execution environments (e.g., Intel SGX) | Trusted isolated execution based on hardware, protect code and data | High cost of trusted hardware, security depends on hardware, limited application scenarios | [115, 121] |
|---|---|---|---|---|
| **Cloud Computing Privacy Protection** | Hybrid cloud architecture and data partitioning | Store and use sensitive data locally, store non-sensitive data in the cloud, reduce leakage risk | Difficult to define data partitioning, high cost of local storage and computation | [101, 103, 108, 115] |
| | Secure data sharing based on proxy re-encryption and attribute-based encryption | Authorization delegation and ciphertext conversion without user interaction, fine-grained access control | High computation overhead, complex key management, limited scalability | [100, 110] |
| | Post-quantum technologies such as quantum-resistant cryptography, quantum key distribution, quantum random number generation | Resist future threats of quantum computer cracking, physically unclonable, true randomness | Poor practicality, extremely high cost, insufficient testing and verification | [25, 120-122] |
| **Data De-identification** | Traditional data publishing privacy models such as k-anonymity and l-diversity | Reduce linkage attack risk, retain certain data utility | High data distortion, weak resistance to background knowledge attacks | [123, 124] |
| | Differential privacy noise addition | Quantified strict privacy protection, resist various attacks | Difficult to determine noise scale, large utility loss, insufficient robustness | [103, 119] |
| | Synthetic data generation | Do not expose original data, retain data feature distribution | Quality depends on the model, poor specificity, high computation overhead | [125, 126] |
| | Federated learning privacy | No need to centralize original data, participants are controllable | Performance depends on data distribution, difficult to prevent | [99, 109, 111, 112, 118, 127] |

| | | | inference attacks, high communication overhead | |
|---|---|---|---|---|
| | Secure multi-party computation | Do not disclose individual privacy, accurately compute aggregate results | High computation and communication complexity, poor scalability, participant collusion problem | [97, 114, 118] |
| **Cryptographic Privacy Protection** | Homomorphic encryption | Ciphertext computation, multi-party collaboration, privacy-preserving queries and machine learning | Low computation efficiency, difficult key management, poor practicality | [97, 98, 102, 106, 107, 109, 113, 114, 116] |
| | Secure multi-party computation | Joint statistical analysis, no need to share original data | High communication overhead, complex protocols, participant collusion problem | [97, 114, 118] |
| | Zero-knowledge proof | Prove statements without providing original information | Low proof generation and verification efficiency, heavy computation burden on the prover | [84, 85] |
| | Functional encryption | Authorized restricted key can decrypt function value, ciphertext can be computed and updated | Complex construction, severe ciphertext expansion, limited function classes | [90] |
| | Model aggregation under differential privacy | Quantified strict privacy protection, prevent inference and membership inference attacks | Difficult to balance noise with utility and privacy, affect model performance, high communication overhead | [111, 119] |
| | Gradient aggregation under secure multi-party computation | Accurate and unbiased global gradient, individual gradients are not disclosed | High computation overhead, multiple interaction rounds, low efficiency, participant collusion problem | [118] |

| **Federated Learning Privacy Protection** | Protect local model training with homomorphic encryption | No need to disclose local data and models, verifiable | Huge encryption and decryption computation overhead, reduced model accuracy | [109, 127] |
|---|---|---|---|---|
| | Blockchain incentive and verification of federated learning process | Participation incentive, proof of contribution, behavior verification, model auditing | Limited throughput, frequent interactions, introduce new security risks | [99, 127] |

This pie chart illustrates the distribution of six privacy protection technologies across 52 research papers. Blockchain technology leads with 17 papers, representing 32.7%, indicating its high relevance in privacy protection. Cloud computing follows with 15 papers (28.8%), highlighting its advantages in data storage and processing. Data de-identification appears in 14 papers (26.9%), underscoring its critical role in privacy defense. Access control and Cryptography each have 13 papers (25.0%), emphasizing their importance in permission management and data protection. Finally, Federated Learning appears in 6 papers (11.5%); although its share is smaller, its potential in distributed privacy protection is gaining attention. This chart clearly reflects the current research focus on different privacy protection technologies in the literature. As shown in Figure 4-2 below.

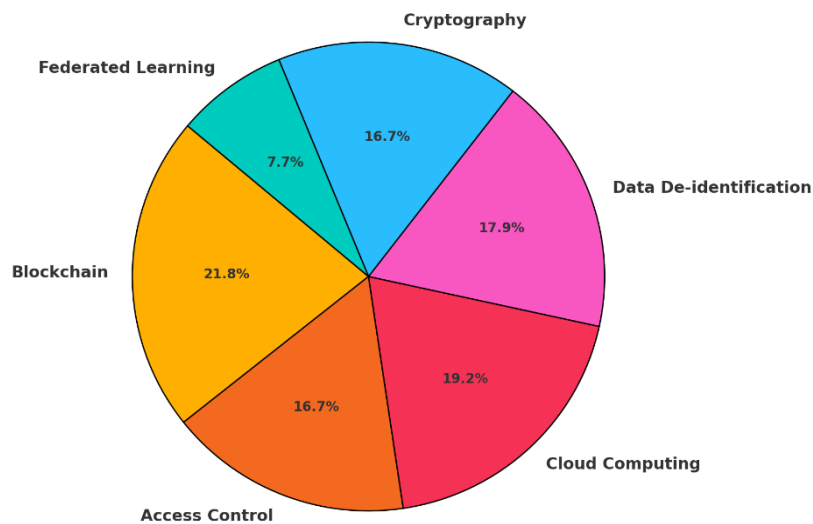

**Fig. 4-2. Frequency Of Privacy Protection Technologies in 52 Research Papers**

### 4.0.3 What are the legal and regulatory studies on health big data in the existing literature? (RQ3)

To further explore the similarities and differences in health data privacy and security across different regions, this study conducted a systematic review and comparative analysis of 12 relevant articles. The European Union, through GDPR [41, 129-131], and China, through PIPL[132], have taken significant steps in establishing specialized legislation that emphasizes individual rights and data protection principles. In contrast, the United States relies on HIPAA [133-135], but this act has limitations in addressing privacy risks associated with de-identified and non-clinical health data, while the United Kingdom lacks a specialized legal framework for research purposes involving primary care electronic health records[136].

Despite these differences, all regions generally recognize the importance of protecting patient privacy, promoting data security, and supporting scientific research. A global perspective[137-139] highlights the need for governments to strengthen privacy protection systems, clarify the responsibilities of data controllers, and address challenges posed by advanced analytical technologies. The EU prioritizes data minimization and restrictions on cross-border data transfers [41, 129-131]; the U.S. focuses on enhancing the interoperability and security of electronic health records[133-135]; China emphasizes the principles of legality, legitimacy, and necessity in data processing [132]; and the UK seeks to balance research uses with patient privacy [136].

In summary, the comparative analysis demonstrates ongoing global efforts to address the complex landscape of health data protection, with a shared goal of promoting responsible data use while safeguarding individual rights and privacy.

By extracting key information, the following comparison table was constructed to present the main characteristics and differences in regulatory backgrounds, legislative trends, and key concerns in each region.: as shown in table 4-3.

**Table 3.** A Cross-Regional Comparison of Health Data Privacy and Security

| Country/Region | Regulatory Background and Legislative Trends | Key Privacy and Security Concerns | Relevant Literature |
|---|---|---|---|
| European Union | GDPR is the EU's latest comprehensive data protection regulation, balancing the promotion of scientific research and privacy protection. Recent legislative amendments avoid excessive restrictions on medical research data sharing. | Emphasizes individual ownership and control over medical and health data. Focuses on data minimization, anonymization, and other security measures. Cross-border transfers of sensitive data must comply with GDPR requirements. | [41, 129-131] |
| United States | HIPAA is the core legislation for medical and health privacy protection, but a large amount of data falls outside its scope. Patients' privacy concerns affect healthcare-seeking behavior, requiring improved legislation to address big data privacy risks. | PIPL reinforces the legality of personal information processing and enhances sensitive information protection. Users' attention to health code privacy policies, trust, and acceptance of purposes influence their perception of privacy protection. | [133-135] |
| China | PIPL is the first dedicated legislation comprehensively regulating personal information processing. Health code apps are widely used, and their privacy protection is a key concern. | PIPL reinforces the legality of personal information processing and enhances sensitive information protection. Users' attention to health code privacy policies, trust, and acceptance of purposes influence their perception of privacy protection. | [132] |
| United Kingdom | Lacks dedicated legal and policy frameworks for regulating the use of primary care electronic health records in research Needs to leverage data research value while protecting privacy. | Data de-identification, secure transmission and storage, and transparency are crucial. Strengthening public engagement and increasing acceptability of purposes. | [136] |
| Global | Existing laws lag behind and struggle to keep pace with big data technology developments. Calls for establishing physiological data protection rules and clarifying individual ownership. Should open up medical data to prevent privatization issues. | Big data analysis and utilization may harm privacy and lead to discrimination; data controllers' responsibilities should be clarified. Black box algorithms exacerbate bias and unfairness, requiring stronger auditing, accountability, and individual control. Governments should promote medical data open sharing while improving privacy protection and usage regulations. | [137-139] |

This geographical heatmap illustrates the distribution of 12 studies on health data privacy and security. The color intensity and concentration of the heat areas visually represent the research density across different regions. It can be observed that the European Union (approximated by Brussels' coordinates) is the most concentrated area, marked in red, representing four studies. The United States and China follow, both represented with blue hotspots, with three and one study, respectively. The United Kingdom also appears in blue, indicating fewer studies. Additionally, a green hotspot near the equator represents three studies with a global perspective. This map suggests that the European Union and the United States have relatively high attention in health data privacy research. As shown in Figure 4-3 below.

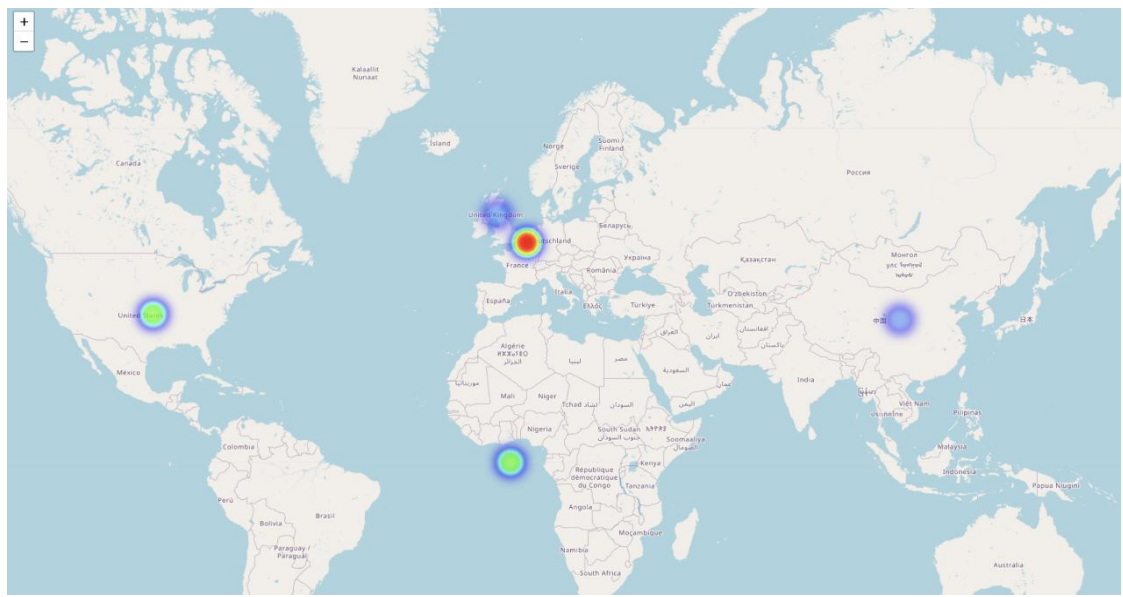

**Fig. 4-3. Global Distribution of Health Data Privacy and Security Research**

**5.0 Challenges and Recommendations**

Through systematic analysis of the selected papers, we have summarized the main challenges in medical big data privacy protection and propose the following solutions and recommendations:

**1. The Conflict between Data Sharing and Privacy Protection**
**Challenge:** While the widespread sharing of medical big data is beneficial for scientific research and medical progress, it may also compromise patient privacy. The literature points out that tech companies collect and control a large amount of medical data, but patients cannot access and share their own data, and researchers have difficulty accessing it, hindering scientific progress[139]. At the same time, overly strict privacy protection regulations may limit the use of data in scientific research[135].
**Suggestions:**
(1) Strengthen patients' control over their personal data, allowing them to actively share data for scientific research and build an open medical data network [139]. The latest data protection legislation in the EU allows data sharing for research purposes under certain circumstances, which is worth learning from [131].

(2) Develop privacy-preserving technologies such as federated learning and secure multi-party computation to achieve data sharing while protecting privacy [87, 104, 112, 118]. Research from institutions like the University of California, Irvine has shown that federated learning can achieve results comparable to centralized models without sharing raw data [112].

(3) Improve laws and regulations, clearly define the scope and requirements of data sharing, and balance innovation needs and privacy protection [136, 137]. We can refer to regulations such as the EU's GDPR, which exempts research use restrictions while strengthening the protection of patients' rights such as informed consent [130].

**2. Privacy Risks of Health Data in Non-Medical Scenarios**

**Challenge:** Medical data has gone beyond the scope of medical institutions. Social media and smart devices are collecting a large amount of personal health information, but lack privacy protection [134]. Data brokerage companies collect and integrate this data for commercial purposes, potentially violating privacy and causing discrimination [134]. Physiological data should also be considered confidential, but currently lacks regulation [138].

**Suggestions:**

(1) Expand the scope of health privacy protection, bring health data outside of medical scenarios under regulation, such as developing targeted regulations and clarifying data collection and usage rules[134, 137]. The EU's GDPR has already defined health data very broadly[130].

(2) Restrict data collection and use for commercial purposes, strengthen user informed consent, give users more control rights, and reduce the risk of data abuse[132, 134]. A survey in Wuhan and Hangzhou, China found that 60% of respondents were very concerned about the privacy policy of health code apps [132].

(3) Develop user-friendly privacy protection technologies and increase user privacy awareness, such as the "Do Not Track" function in browsers and fine-grained access control for personal health records[82, 89, 93].

**3. Data Quality and Management Issues in Medical Data**

**Challenge:** The data quality of electronic health record (EHR) systems varies, interoperability is poor, making it difficult for doctors to use and researchers to apply [135]. There is a lack of comprehensive guidance on various aspects of data management, such as how to collect, store, and access data [136].

**Suggestions:**

(1) Improve data standards for EHR systems, strengthen cooperation among vendors, and enhance system compatibility and interoperability[135, 136]. The TEFCA program in the US aims to achieve interconnection between different systems [136].

(2) Develop best practice guidelines for medical data lifecycle management, covering best practices for collection, transmission, storage, analysis, access and other aspects, and provide supporting tools such as assessment templates[136]. Primary care research in the UK provides a 6-step model for using EHR data [136].

(3) Utilize emerging technologies to improve data quality and management, such as

blockchain technology to ensure data traceability and immutability[86, 90, 97, 104]; artificial intelligence and natural language processing can simplify data entry and reduce the burden on doctors [135].

**4. Privacy Protection Policies and Mechanisms Need Improvement**

**Challenge:** Existing privacy regulations such as HIPAA have limited coverage [134] and are difficult to address the challenges of medical privacy protection in the era of big data. There is a lack of targeted industry guidelines, standards, and mechanisms [129, 136]. Enforcement is weak, the cost of violations is low, making it difficult to truly protect privacy rights [137].

**Suggestions:**

(1) Accelerate the construction of a health privacy protection legal system and expand the breadth and depth of legal coverage. For example, we can learn from GDPR to formulate comprehensive laws for medical big data, and improve supporting industry guidelines and ethical review mechanisms[129, 130, 137].

(2) Strengthen the privacy protection capacity building of medical institutions, incorporate privacy protection into the performance evaluation system, and increase the cost of violations. Refer to the US ONC's Guide to Privacy and Security of Electronic Health Information[136] and the EU GDPR's high penalty provisions [130].

(3) Establish a dedicated privacy protection regulatory agency and strengthen enforcement. Establish patient privacy protection officers to supervise implementation and promptly discover and deal with violations [137].

**5. Barriers to Cross-Border Data Flows**

**Challenge:** The differences in privacy protection policies between countries and regions are large, creating barriers to cross-border medical cooperation and data sharing [130]. The competition for data sovereignty among countries is becoming increasingly fierce [137].

**Suggestions:**

(1) Strengthen international dialogue and cooperation, and promote convergence of privacy protection policies. For example, participate in relevant initiatives of organizations such as OECD and APEC, develop guidelines for cross-border medical data flows, and build mutual trust mechanisms[130, 137].

(2) Establish a mutual recognition mechanism for data ethics review to reduce the time and cost of repeated reviews. Such as the EU's "one-stop approval, multi-site recognition" mechanism [130].

(3) Develop trusted data circulation technology and protect privacy at the source. For example, adopt distributed analysis technologies such as federated learning so that parties can model locally without physically transferring data[87, 104, 112]; apply cryptographic techniques such as encrypted computing and secure multi-party computation to achieve data sharing "as is"[97, 104, 118].

**6. Addressing Privacy Threats from New Technologies like Quantum Computing**

**Challenge:** The development of quantum computing poses a threat to traditional cryptography and may break medical data encryption mechanisms. How to deal with privacy protection in the post-quantum era is a big challenge[25, 120, 122].

**Suggestions:**

(1) Accelerate the research, development and application of new cryptographic technologies such as post-quantum cryptography in the medical field. Use quantum-resistant algorithms such as quantum key distribution and quantum digital signatures to encrypt and authenticate medical data[121, 122]. Due to the computational complexity of post-quantum cryptography, it can first be used for critical, long-term preserved medical data [25].

(2) Combine privacy protection technology with blockchain technology to enhance system robustness using decentralized networks. Adopt quantum blockchain, combining quantum cryptography and blockchain consensus mechanisms, to achieve quantum-resistant privacy protection and data auditing[121, 122].

(3) Establish global cooperation to jointly address quantum threats. Develop international standards, strengthen cooperative research on post-quantum cryptography, and form a consensus on responses [25].

## 7. Lack of Privacy-Preserving Primitives and Systems for Healthcare

**Challenge:** Existing privacy protection primitives and systems are mainly designed for general scenarios and are difficult to apply directly to the healthcare industry. How to develop efficient, easy-to-use privacy protection technologies that meet medical needs is a major challenge[98, 99, 107, 110].

**Suggestions:**

(1) Conduct a survey of medical privacy protection needs as a basis for technology research and development. Comprehensively understand users' privacy concerns and data usage needs in different medical scenarios to guide the design of privacy protection systems[109, 113].

(2) Develop high-performance medical-specific cryptographic components. Optimize cryptographic algorithms, reduce encryption and decryption overhead, and minimize the impact on medical operations[26, 96, 116]; design medical-specific key management and access control mechanisms[88, 94, 99].

(3) Provide easy-to-use privacy protection systems. Adopt intuitive policy configuration tools to lower the threshold for using privacy protection; develop privacy protection systems that seamlessly integrate with medical information systems [82, 118, 128]; provide supporting security situational awareness and auditing functions to help medical institutions continuously evaluate and improve privacy protection[101, 114].

In conclusion, healthcare big data privacy protection requires a multi-stakeholder effort across technical, regulatory, and managerial dimensions to balance data utilization with privacy protection, fostering a favorable environment for the development and application of healthcare big data for public health benefits.

## 6.0 Conceptual Framework

Here I present a hierarchical privacy protection framework inspired by biological immune systems. The architecture consists of four interconnected layers that work synergistically to provide autonomous, adaptive, and quantum-resistant privacy protection. a, The Bio-Inspired Privacy Core mimics the human immune system's [39]self-adaptive defense mechanism, incorporating a Data Perception Layer that continuously monitors system state, an Immune Memory Bank that archives defense patterns, an Adaptive Evolution Engine that optimizes defense strategies, and a Dynamic Defense Decision module that executes real-time protective actions. b, The Multi-Agent Privacy Guards layer implements distributed defense through specialized privacy sentinel agents (n ≥ 3), each focusing on specific threat patterns while maintaining cross-agent communication for coordinated responses. c, The Quantum-Safe Security Layer ensures post-quantum security through three complementary mechanisms: post-quantum homomorphic encryption[25] enabling computation on encrypted data, zero-knowledge proofs[24] for trustless verification, and quantum key distribution for unbreakable key exchange. d, The Federated Intelligence Network layer enables privacy-preserved distributed learning across multiple nodes (minimum 3 nodes shown), coordinated through a blockchain-based[30] consensus mechanism and automated by smart contracts. Arrows indicate data flow and control signals between components. As shown in Figure 6-1 below.

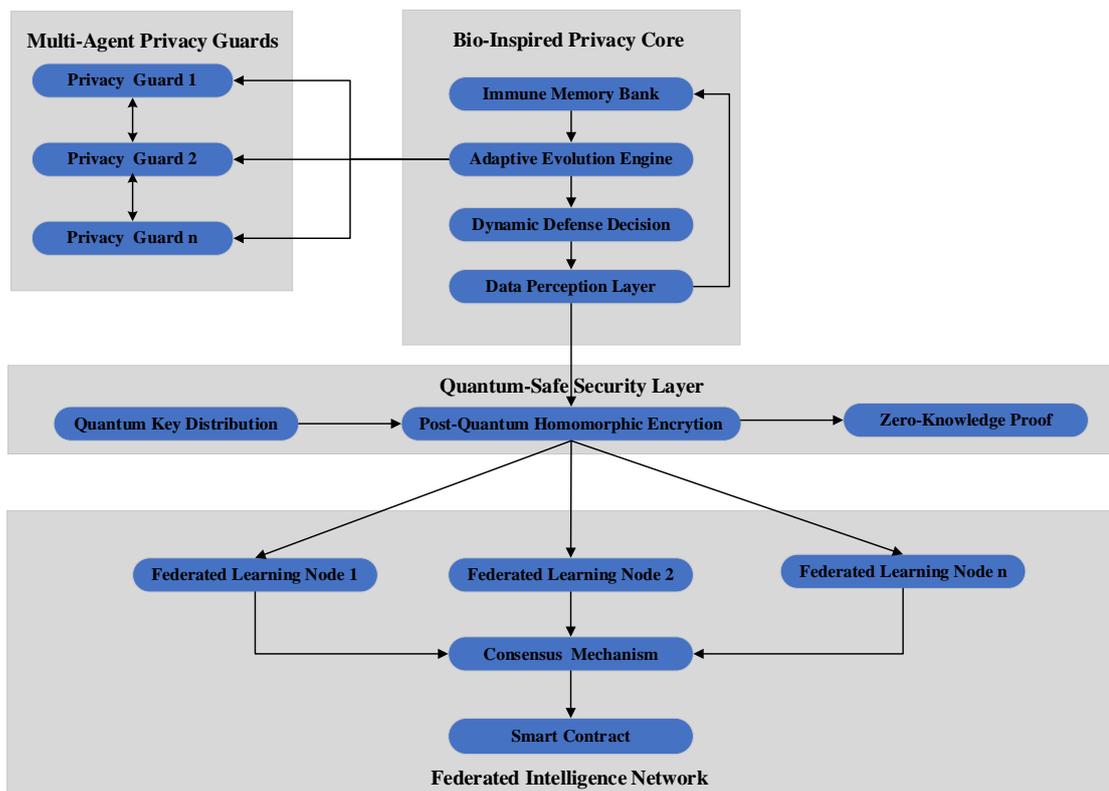

**Fig. 6-1 Biologically-Inspired Adaptive Framework for Health Data Privacy Protection**

## 7.0 Conclusion

This systematic review aims to understand the scope of privacy-preserving technologies in the field of health big data. To achieve this objective, we conducted a systematic literature review of eight authoritative digital databases following a specific protocol to identify suitable research articles. The findings emphasize the crucial role of advanced technologies and regulatory frameworks in protecting health big data privacy. By integrating the limitations of existing research, recommendations, and knowledge gaps identified in this review, we proposed a comprehensive framework and suggested future research directions. These research directions should focus on refining relevant technologies and ensuring their compliance with evolving legal standards.